%% file: band_parallelization.tex
\documentclass[final,5p,times,letterpaper,twocolumn]{elsarticle}
\usepackage[utf8]{inputenc}
\usepackage{amssymb}
\usepackage{amsmath}
\usepackage{listings}
\usepackage{hyperref}
\usepackage{braket}
\usepackage{enumitem}
\include{macros}

\journal{Computer Physics Communications}

\begin{document}
\begin{frontmatter}
\title{
Enhancement of DFT-calculations at Petascale: Nuclear Magnetic Resonance, Hybrid Density Functional Theory and Car-Parrinello calculations
}
\author[ichec,ivec,curtin]{Nicola Varini\corref{cor1}\fnref{fm1}}
\author[istm]{Davide Ceresoli\fnref{fm2}}
\author[university-gorica,iom]{Layla Martin-Samos}
\author[ichec,ictp]{Ivan Girotto}
\author[cineca]{Carlo Cavazzoni}
\address[ichec]{ICHEC, 7th Floor Tower Building Trinity Technology
               \& Enterprise Campus, Grand Canal Quay, Dublin 2, Ireland}
\address[ivec]{iVEC, 26 Dick Perry Ave, Kensington WA 6151, Australia}
\address[curtin]{Research and Development, Curtin University, GPO Box U 1987, Perth, WA 6845, Australia}
\address[istm]{CNR Istituto di Scienze e Tecnologie Molecolari (CNR-ISTM),
               c/o Dept. of Chemistry, University of Milan, via Golgi 19,
               20133 Milan, Italy}
\address[iom]{CNR Istituto Officina Molecolare (CNR-IOM),
              c/o SISSA, via Bonomea 235, 34136 Trieste, Italy}
\address[ictp]{International Centre for Theoretical Physics (ICTP),
               Strada Costiera 11, 34014 Trieste, Italy}
\address[cineca]{CINECA, via Magnanelli 6/3, 40033 Casalecchio di Reno, Italy}
\address[university-gorica]{University of Nova Gorica, Materials Research Laboratory, vipavska cesta 11C, 5270 Ajdovscina, Slovenia}
\cortext[cor1]{Corresponding author.}
\fntext[fm1]{E-mail: \url{nicola@ivec.org}}
\fntext[fm2]{E-mail: \url{davide.ceresoli@istm.cnr.it}}
\begin{abstract}

One of the most promising techniques used for studying the electronic properties of materials 
is based on Density Functional Theory (DFT) approach and its extensions. 
 DFT has been widely applied in traditional solid state physics problems where
periodicity and symmetry play a crucial role in reducing the computational workload. 
With growing compute power capability and the development of improved DFT methods, the range of 
potential applications is now including other scientific areas such as Chemistry and Biology. 
However, cross disciplinary combinations of traditional Solid-State Physics, Chemistry and Biology
drastically improve the system complexity while reducing the degree of periodicity and symmetry.  Large simulation cells
containing of hundreds or even thousands of atoms are needed to model these kind of physical systems. 
The treatment of those systems  still remains a computational challenge
even with  modern supercomputers. 
In this paper we describe our work to improve the scalability of 
Quantum ESPRESSO \cite{Quantum-Espresso} for treating very large cells and huge numbers of electrons. 
To this end we have introduced an extra level of parallelism,
over \emph{\emph{electronic bands}}, in three kernels for solving computationally expensive problems:
the Sternheimer equation solver (Nuclear Magnetic Resonance, package QE-GIPAW), the Fock operator builder (electronic ground-state, package PWscf) and most of the Car-Parrinello routines (Car-Parrinello dynamics, package CP). Final benchmarks show our
success in computing the Nuclear Magnetic Response (NMR) chemical shift
of a large biological assembly, the electronic structure of defected
amorphous silica with hybrid exchange-correlation functionals and the equilibrium atomic structure of height Porphyrins anchored to a Carbon Nanotube, on many thousands of CPU cores.
\end{abstract}

\end{frontmatter}

\section{Introduction}
Understanding the details of atomic/molecular structures as well as dynamics of
solid materials and biological systems, is one of the major challenges
confronting physical and chemical science in the early 21st Century. Such
knowledge has direct impact on important issues in our society like: the design, synthesis 
and processing of new either eco-friendly or
high-efficiency materials; development of new energy sources; control
of materials degradation and recycling; design of drugs for specific pathology.  

The complexity of these areas of research is
growing rapidly along with the need for computational tools to deal with
such complexity.  Computer simulation is the only way to
study these large physical systems. However, extremely expensive large-scale
computational resources do not come for free. In addition to the scientific
value of the computer simulation,  scientific codes must show good scalability
and efficiency for access to world-class supercomputing
facilities.

In this respect, parameter-free first principles (aka \emph{ab-initio})
atomistic calculations in the framework of Density Functional Theory
(DFT)~\cite{hk64,ks65} are quite popular models, as they combine
reasonable chemical accuracy with affordable complexity and good
scalability. The current available codes implementing those models
\cite{Quantum-Espresso,ABINIT,SIESTA,CPMD,VASP,QBOX} have enabled computer simulations 
at the frontier of science. 

Computer technology is rapidly changing. 
The trajectory of major computer manufactures of the last few years illustrate that vendors are rapidly
increasing the number of cores to build supercomputing technology,
reducing the clock frequency of each single compute element. The most powerful supercomputers in the world are today equipped with
million of cores capable of scheduling billions of threads concurrently.
Without substantial updating, many legacy codes will not be able to exploit
this massive parallelism. Nevertheless,
such horsepower gives the opportunity to address problems of unprecedented size and complexity.
This paper presents a parallelization strategy, based on the distribution of
electronic band loops, for efficiently scaling calculations
of hybrid-functionals and Nuclear Magnetic Resonance (NMR).
In addition we describe how the same parallelization
strategy impacts in the scalability of the Car-Parrinello computational kernels.

The paper is organized as follows. In Sec.~\ref{sec:related} we briefly
review related work reported by other groups.
In Sec.~\ref{sec:gipaw} we describe the
equations implemented in the QE-GIPAW code to calculate NMR shielding
tensors and in Sec.~\ref{sec:exx} the formulas needed to evaluate the
Fock-exchange operator and energy.  In Sec.~\ref{sec:parallelization_strategy} we
outline the parallelization strategy based on \emph{bands} and $q$-point
distribution. In Sec.~\ref{sec:benchmark} we report and discuss benchmark
results of our parallelization. 
Finally, in the last section we
present our conclusions and perspectives on this parallelization strategy.

\section{Related work}
\label{sec:related}
The idea of distributing computation over \emph{electronic bands}
was pioneered in the '90s by the CASTEP~\cite{hasnip09} code and was found to be
effective on early vector machines (i.e. Cray Y-MP). In the subsequent years,
the improvement of collective communication along with the availability of efficient
FFT libraries, enabled a good scaling of plane wave codes up to few hundreds of
CPUs with a simple two-level parallelization (k-points and plane waves).
Only since very recently, top-level HPC machines have a huge number
of CPU cores (10$^4$--10$^5$) with limited amount of memory. As
the number of atoms in the system increase, the number of k-points can
be reduced, deteriorating the scaling and parallel efficiency of the
codes. At this stage, in order to exploit the large number of CPU cores,
an extra level of parallelism, \emph{electronic bands}, must be introduced.
The iterative diagonalization has been the first kernel to
benefit of this strategy allowing a scalability up to few thousands of cores
for electronic groundstate calculations within traditional exchange-correlation
functionals.
ABINIT implemented
a block preconditioned conjugate gradient algorithm~\cite{bottin08},
while VASP avoids explicit orthogonalization of bands by the RM-DIIS
algorithm. Moreover, the QBOX~\cite{QBOX} code was designed purposely for
BlueGene machines by a clever data distribution and fully distributed
linear algebra (ScaLAPACK~\cite{Scalapack}). Similar schemes have
been implemented in CPMD, BigDFT, CASTEP~\cite{hasnip09} and are being
implemented within different packages of the Quantum ESPRESSO distribution. 
Regarding the Exact Exchange (EXX) kernel, the NWCHEM code 
introduced an extra level of
parallelism over occupied states and reported excellent scaling up
to 2,048 CPUs~\cite{NWCHEMEXX}.
In this work we report scaling results up to many thousands of CPU core,
thanks to the band parallelization on replicated data of two of the most computationally intensive
kernels (linear response and the Fock operator). In addition to this,
we show that every step of the Car-Parrinello algorithm can be fully
distributed (both data and computation), enabling petascale computation.
This parallelization stragtegy allows to fully exploit
the new EU Tier-0 petascale facilities such as CURIE~\cite{Curie}, JUGENE~\cite{Jugene} and FERMI~\cite{Fermi}.

\section{The GIPAW equations for the induced current and NMR shieldings}
\label{sec:gipaw}
The GIPAW method (Gauge Including Projector Augmented Wave) makes it possible
to calculate the current induced by an infinitesimal external field,
hence the NMR shielding tensors, in a periodic system by means
of linear response. The GIPAW method is an extension of the PAW
method~\cite{bloechl94} for reconstructing all-electron
wavefunction and expectation values from a pseudopotential calculation.
This is essential to compute accurately the response of the valence
electrons in regions near the nuclei, which determines the NMR shielding.

The GIPAW method was formulated by Mauri~\cite{mauri96} and
Pickard~\cite{pickard01,pickard03,yates07} and we refer the reader
to the original papers for the detailed derivation of the method.
In the following, we  show only the resulting set of equations for the
induced current and  outline the flow of the code.

We will start by noting that the GIPAW transformation of a quantum
mechanical operator $\Ocal$ (eq.~17 of Ref.~\cite{pickard01}), gives
rise to two kind of terms. The first is the operator $\Ocal$ itself,
(which we call \emph{bare} term), acting on the valence wavefunctions
all over the space.  The second is a non-local operator acting only
inside spherical augmentation regions centered around each atom. The
non-local projectors enjoy the property of vanishing beyond a cutoff
radius, equal to the cutoff radius of the pseudopotential. This term is
called \emph{reconstruction} term and is not computationally expensive,
except for systems with more than 2000 atoms.  For larger systems,
work is in progress to block-distribute the non-local projectors and
use the ScaLAPACK~\cite{Scalapack} library to perform linear algebra operations.

Therefore, in the following we will focus on the evaluation of the
\emph{bare} contribution to the induced current and to the magnetic susceptibility
of the system. The induced current is defined as
\begin{multline}
  \jvec_\mathrm{bare}^{(1)}(\rvec',q) = \frac{1}{c} \sum_{k} w_k 
  \sum_{\qvec = \pm q\hat{x}, \pm q\hat{y}, \pm q\hat{z}}\frac{1}{2 q}
  \sum_{n\in\mathrm{occ}}
  \mathrm{Re} \left[ \frac{2}{i} \right.\\
  \quad\left.\braket{u_{n\kvec}|\Jvec^\mathrm{para}_{\kvec,\kvec+\qvec}(\rvec')\,
  \Gcal_{\kvec+\qvec}(\epsilon_{n\kvec})\, \Bvec \times \hat{\qvec} \cdot 
  \vvec_{\kvec+\qvec,\kvec} | u_{n\kvec} }\right]
  \label{eq:j}
\end{multline}
where the first sum is over k-points with weight $w_k$. The second sum
is over a star of q-points, where $q\ll1$ is the inverse wavelength of
the external magnetic field and $\hat{\qvec}$ is a unit vector. The third
sum is over the occupied bands, $u_{nk}$ is the valence eigenstate at
k-point $k$ and band index $n$.
The bracket is evaluated from right to left, by applying first the
non-local velocity operator
\beq
  \vvec_{\kvec,\kvec'} = -i\nabla + \kvec' + \frac{1}{i}
  \left[\rvec,V^\mathrm{nl}_{\kvec,\kvec'}\right]
  \label{eq:velocity}
\eeq
to $\ket{u_{nk}}$, then by applying the Green's function, formally defined as
\beq
  \Gcal_{\kvec+\qvec}(\epsilon) = \left( \Hcal_{\kvec+\qvec,\qvec} - 
    \epsilon\right)^{-1}
   \label{eq:greenfunction}
\eeq
where $\Hcal_{\kvec',\kvec} = e^{-i\kvec'\cdot\rvec}\Hcal e^{i\kvec\cdot\rvec}$
is the periodic Kohn-Sham Hamiltonian.
Then, we apply the paramagnetic current operator to the left to $\bra{u_{nk}}$
\beq
  \Jvec^\mathrm{para}_{\kvec,\kvec'} = -\frac{1}{2}\left[
  (-i\nabla+\kvec)\ket{\rvec'}\bra{\rvec'} +
  \ket{\rvec'}\bra{\rvec'}(-i\nabla+\kvec') \right]
  \label{eq:jpara}
\eeq
Finally the braket product is evaluated and summed to obtain the current
field. The induced magnetic field is obtained easily by the Biot-Savart
law, after Fourier-transforming to reciprocal space:
\beq
  \Bvec_\mathrm{bare}^{(1)} = \frac{4\pi}{c} \frac{i\,
  \Gvec\times\jvec_\mathrm{bare}^{(1)}(\Gvec)}{G^2}
  \label{eq:biot}
\eeq
This procedure is repeated  with $\Bvec$ along each of the three Cartesian
directions, and the \emph{bare} NMR shielding tensor is obtained by
evaluating the induced magnetic field response at each nuclear coordinate.
The $\Gvec=0$ term in eq.~\ref{eq:biot} depends on the macroscopic shape
of the sample and the magnetic susceptibility tensor $\chi_\mathrm{bare}$.
The expression of $\chi_\mathrm{bare}$ is very similar to eq.~\ref{eq:j}:
\begin{multline}
  \tensor{Q}(q) = -\frac{1}{c^2} \sum_{k} w_k\,\, \frac{1}{\Omega}\,
  \sum_{\qvec=q\hat{x},q\hat{y},q\hat{z}}\,\,
  \sum_{n\in\mathrm{occ}}
  \mathrm{Re} \left[ \frac{1}{i} \right.\\
  \quad\left.\braket{u_{n\kvec}|\hat{\qvec} \times (-i\nabla+\kvec)\,
  \Gcal_{\kvec+\qvec}(\epsilon_{n\kvec})\, \Bvec \times \hat{\qvec} \cdot 
  \vvec_{\kvec+\qvec,\kvec} | u_{n\kvec} }\right]
  \label{eq:Q}
\end{multline}
where $\Omega$ is the cell volume and
\beqa
  \tensor{\chi}_\mathrm{bare} &=&
  \frac{\tensor{F}(q) - 2\tensor{F}(0) + \tensor{F}(-q)}{q^2}\nn
  F_{ij}(q) &=& (2-\delta_{ij})\,Q_{ij}(q) \hspace{1cm} i,j \in x,y,z
\eeqa
Since the only difference between eqs.~\ref{eq:j} and \ref{eq:Q} is the
operator to the left of the Green's function, the two formulas are evaluated
back-to-back in the same loop.

Finally, the NMR shielding tensor $\tensor{\sigma}(\Rvec)$ is obtained by
adding the bare and reconstruction contributions of the induced magnetic
field, evaluated at each nuclear position $\Rvec$
\beq
  \tensor{\sigma}(\Rvec) = \left.
  -\frac{\partial\Bvec^{(1)}(\rvec)}{\partial\Bvec} \right|_{\rvec=\Rvec}
\eeq
\subsection{The Sternheimer equation}
The most time-consuming operation in GIPAW is applying the Green's
function (eq.~\ref{eq:greenfunction}) to a generic ket $\ket{w_{n\kvec}}$
\beq
  \ket{g_{n\kvec+\qvec}} = \Gcal_{\kvec+\qvec}(\epsilon_{n\kvec})\ket{w_{n\kvec}}
\eeq   
To avoid inversion of large matrices or, equivalently, summation over all empty
states, we solve instead the Sternheimer equation
\beq
  \left( \Hcal_{\kvec+\qvec,\kvec} - \epsilon_{n\kvec} + 
  \alpha P_{\kvec+\qvec,\kvec} \right) \ket{g_{n\kvec+\qvec}} =
  - Q_{\kvec+\qvec,\kvec} \ket{w_{n\kvec}}
  \label{eq:sternheimer}
\eeq
where $P_{\kvec+\qvec,\kvec}=\sum_{n\in\mathrm{occ}}\ket{u_{n\kvec+\qvec}}
\bra{u_{n\kvec}}$ is the projector over the occupied manifold and
$Q_{\kvec+\qvec,\kvec}=1-P_{\kvec+\qvec,\kvec}$ is the projector over the
empty states. $\alpha$ is chosen as twice the bandwidth of the occupied
states in order to make the \emph{lhs} positive definite.

Eq.~\ref{eq:sternheimer} constitutes a set of $N$ independent linear
equations, where $N$ is the number of occupied states and is solved for
$\ket{g_{n\kvec+\qvec}}$ by the conjugate gradient (CG) method.  In order
to take advantage of level-3 BLAS operations, the CG update is performed
initially on all \emph{\emph{electronic bands}}. Then, starting from the second
iteration, the \emph{\emph{electronic bands}} are divided into two groups, each occupying
a contiguous memory area: converged and not converged. A band is converged
when the residual falls below a threshold ($10^{-7}$~Rydberg). Therefore,
we update only the bands which have not yet converged. Before returning
from the subroutine, the bands are sorted according to the original
band index $n$.

\subsection{Outline of the GIPAW code}
Before running an NMR calculation with the QE-GIPAW code~\cite{GIPAW},
it is necessary to run a self-consistent calculation (SCF) with the
PW code on a possibly relaxed structure, in order to obtain the ground
state wavefunction $\ket{u_{n\kvec}}$.  The QE-GIPAW code reads from disk
the wavefunctions and the electronic density generated by PW and computes
the induced current and the magnetic susceptibility.

After an initialization phase, the QE-GIPAW code runs as follows:
\begin{enumerate}
\item Loop over $k$-points
  \begin{enumerate}[label=\arabic{*}.]
  \item Read $\ket{u_{n\kvec}}$ from disk, all bands
  \item Set $q=0$ and for each $n$ and compute
  $\vvec_{\kvec,\kvec}\ket{u_{n\kvec}}$,\\
  $\Gcal_{\kvec,\kvec}(\epsilon_{n\kvec})\vvec_{\kvec,\kvec}\ket{u_{n\kvec}}$,
  and $(-i\nabla+\kvec)\ket{u_{n\kvec}}$
  \item Compute $Q(0)$ according to eq.~\ref{eq:Q}
  \item Loop over the star of $q$-points: $\qvec=\pm q\hat{x},\pm q\hat{y},\pm q\hat{z}$
    \begin{enumerate}[label=\arabic{*}.]
    \item Diagonalize the KS Hamiltonian at $\kvec+\qvec$
    \item For each $n$ compute $\vvec_{\kvec+\qvec,\kvec}\ket{u_{n\kvec}}$,\\
    $\Gcal_{\kvec+\qvec,\kvec}(\epsilon_{n\kvec})\vvec_{\kvec+\qvec,\kvec}
    \ket{u_{n\kvec}}$, and $(-i\nabla+\kvec+\qvec)\ket{u_{n\kvec}}$
    \item Compute $Q(q)$ and $\jvec_\mathrm{bare}^{(1)}(\rvec',q)$
    \end{enumerate}
  \end{enumerate}
\item Parallel execution only: reduce (parallel sum) $Q$ and
$\jvec_\mathrm{bare}^{(1)}$ over $k$-points and planewaves.
\item Solve the Biot-Savart equation (eq.~\ref{eq:biot}), evaluate the
induced magnetic field at each nuclear coordinates, output the NMR
shielding tensors and terminate.
\end{enumerate}
All \emph{reconstruction} terms are evaluated after step~1.3
and step~1.4.3, with little computational cost. When employing
ultrasoft~\cite{uspp} of PAW pseudopotential, there is an additional
evaluation of the Green's function for each $k$ and $q$. The details
can be found in Ref.~\cite{yates07}.

Thus the flowchart of the code is based on three nested loops: over
$k$-points, $q$-star and bands. Each term can be calculated independently
and we anticipate that in order to run efficiently at the petaflop scale,
we have distributed the calculation of every individual term on all the
CPUs. Parallel communication is performed only at the end of the three
loops and consumes little time. The only bottleneck is diagonalization
step at $\kvec+\qvec$. We use the wavefunctions calculated at $\kvec$
as starting vector in order to reduce the number of Davidson or
CG~\cite{Quantum-Espresso} iterations. Unfortunately the Davidson
and CG algorithms cannot be parallelized easily over bands, because
of the Grahm-Schmidt orthogonalization.  We are currently seeking a
diagonalization method which does not require and orthogonalization step,
such as the RM-DIIS method~\cite{VASP}.

Moreover, the explicit diagonalization can be avoided for very large
simulation cells by choosing the magnitude of $q$ equal to the first
reciprocal lattice vector, i.e. $q = 2\pi/a$, where $a$ is the lattice
spacing of a cubic supercell. In fact, by the Bloch theorem, the
wavefunctions at $\kvec+\qvec$ are given simply by:
\beq
  \ket{u_{n,\kvec+\hat{\qvec}2\pi/a}} = e^{i\hat{\qvec}2\pi/a\cdot\rvec}
  \ket{u_{n\kvec}}
\eeq
This method (which we call \emph{commensurate}) provides an enormous
speed-up and scaling for very large systems, but it may worsen the
accuracy of the calculated NMR chemical shift. We are currently testing
this method and results will be reported in a future paper.

\section{The Exact Exchange (EXX) within the planewave method}
\label{sec:exx}
After work of A. Becke in 1993, ``A new mixing of Hartree-Fock
and local density-functional theories''~\cite{becke93},  it is nowadays
becoming very common to include a fraction of Exact eXchange (formally
Fock exchange) in Density Functional calculations. Exchange-correlation
functionals with this fraction are called Hybrid-Functionals
(HFs).  The use of HFs enables at least a partial correction to the
self-interaction error and the inclusion of a non-local contribution to
the Hamiltonian.  In other words, the inclusion of an EXX fraction,
with respect to traditional DFT functionals, Local Density Approximation
(LDA) or Generalized Gradient Approximation (GGA)~\cite{PBE}, is mainly
used to improve the agreement  to experiments of the band
gaps and the energetics of small molecules and solids.

The EXX energy contribution is defined as
\beq
E_{x} = -\frac{1}{2} \sum_{\kvec} \sum_{\qvec} \sum_{i,j}^{occ}\int d\rvec d\rvec'
        \frac{u^*_{i\kvec}(r)\,u_{j\kvec-\qvec}(r)\,
              u_{j\kvec-\qvec}^*(r')\,u_{i\kvec}(r')}{|\rvec-\rvec'|}
\eeq
which in reciprocal space  reads
\beq
E_{x} = - \frac{1}{2}\sum_k \sum_{i,j}^{occ}\sum_{\Gvec} \sum_{\qvec}
        \frac{M^*_{i\kvec,j\kvec-\qvec}(G)\,M_{i\kvec,j\kvec-\qvec}(\Gvec')}
        {|\Gvec+\qvec|^2}
\eeq
where $M_{i\kvec,j\kvec-\qvec}(\Gvec) =
\mathrm{FT}\left[u_{j\kvec-\qvec}^*(r) u_{i\kvec}(r)\right]$ is the
Fourier Transform of the products of two Bloch wavefunctions. These
quantities are usually called ``overlap charge densities''.  In
plane-wave-based codes, the overlap charge densities are calculated
via Fast Fourier Transform (FFT).  From the reciprocal-space
representation the wave-functions are Fourier transformed to real-space,
 the overlap product is calculated and then the result is
transformed back to reciprocal space where the sums over occupied
states, $\qvec$, $\kvec$, and $\Gvec$ are performed~\cite{chawla98,gibson06}.

With the \texttt{PWscf} code of the Quantum ESPRESSO distribution, the
convergence of the Khon-Sham (KS) Hamiltonian containing EXX fraction
is achieved through two nested do loops. The inner loop converges the KS
equations at fixed EXX potential while the outer updates the EXX potential
($V_x$) and checks the overall accuracy. Before adding the first guess
for $V_x$, \texttt{PWscf} makes a first full-self-consistent calculation
with a traditional non-hybrid exchange-correlation functional. The
EXX potential projected on an electronic state $\ket{u_{i\kvec}}$ is
calculated as
\beq
V_{x}\ket{u_{i\kvec}} = -\sum_{j}^{occ}\sum_{\Gvec} \sum_{\qvec}
        \frac{M^*_{i\kvec,j\kvec-\qvec}(G)M_{i\kvec,j\kvec-\qvec}(\Gvec')}
        {|\Gvec+\qvec|^2}
\eeq
It is easy to see that the distribution of the computational workload
at the level of the sum over $j$ is very convenient as it just implies
a reduction at the end of the loops.

\begin{figure*}[!th]
\centering
\includegraphics[width=5in]{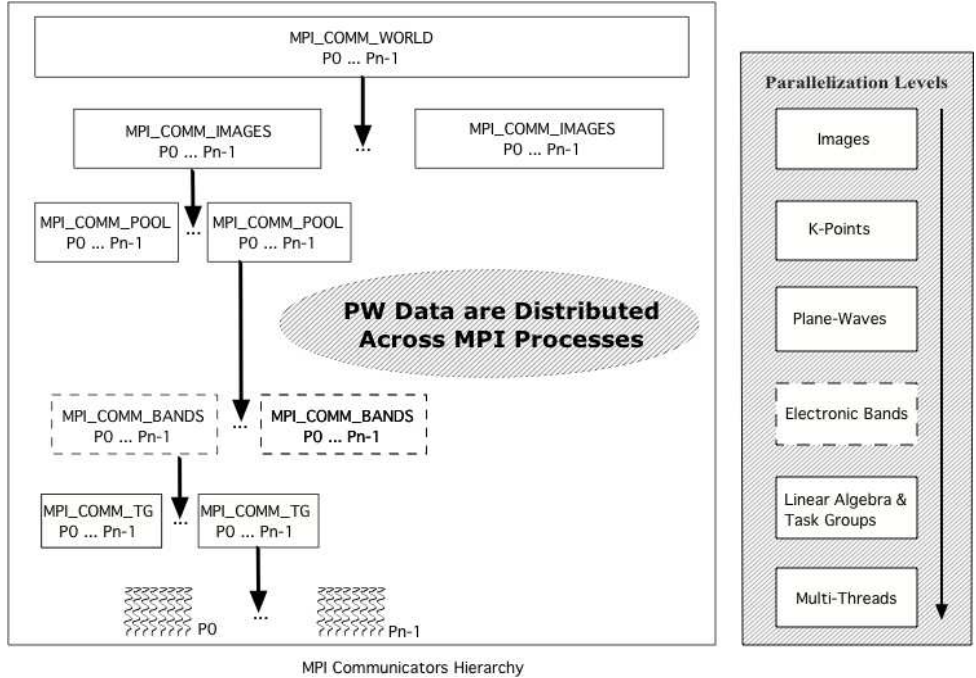}
\caption{The picture stands for the levels of parallelism implemented within 
the Quantum ESPRESSO software packages. On the right hand side the hierarchy 
is logically represented from the higher to the deeper parallelism, expressed 
using multi-threading on top of MPI distribution. The left hand side shows 
how this hierarchy is mapped on MPI groups of processes (or communicators). 
From the top to the bottom four of these levels are divided in smaller 
groups from previous level (each black arrow stands for a splitting stage), 
starting from MPI COMM WORLD which represents all the MPI processes. 
For each sub-groups processes are identified from 0 to n - 1 where n is the 
number of processes for a given MPI communicator. At the last stage each process 
is no longer considered as a member of a set of processes (MPI communicator) but 
a single entity from which threads are created.}
\label{fig:parallelization}
\end{figure*}

Note however that in the current Quantum ESPRESSO implementation, hybrid functionals
cannot be used to compute linear response quantities (i.e. NMR) because
one should implement and solve simultaneously the coupled Hartree-Fock
equations~\cite{CPHF}.  On the contrary, in pure DFT (no EXX) the
zeroth order (Kohn-Sham) and first order (Sternheimer) equations are
fully decoupled.

\section{Parallelization strategy}
\label{sec:parallelization_strategy}
Historically, the Quantum ESPRESSO distribution has been designed and used
for modeling condensed-matter problems where the physical system can be
mapped into a periodically repeated primitive cell. In these cases, the
number of electrons (and consequently the number of \emph{\emph{electronic bands}})
is a relatively small number while the number of k-points (Brillouin
zone sampling) is usually high. As a result, the k-points distribution
across processes is the natural method of parallelization. Especially
if considering that the DFT Hamiltonian is diagonal in k and the only
source of communication arises from a sum in the calculation of the total
energy. However, the boom of new organic and hybrid based electronics
and optoelectronics materials (originated by the need of replacing silicon-based
technologies) along with the fast evolution of supercomputers present
brand-new scenarios extending the boundaries among condensed-matter,
chemistry and biology. Most of the new topics involve systems with a
huge number of atoms (subsequently, a huge number of electrons) and
either low-symmetry (i.e., surfaces or wires) or no-symmetry at all
(i.e., biological molecules).

\begin{figure*}[!th]
\centering
\includegraphics[width=5in]{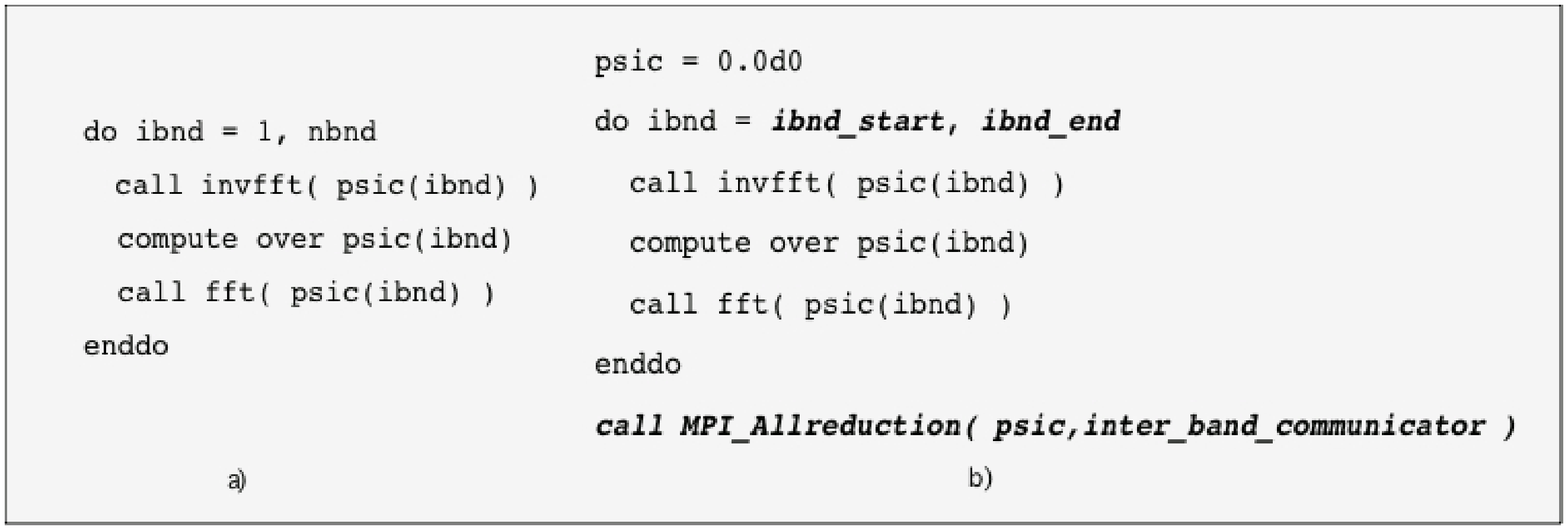}
\caption{Code sections from QE-GIPAW code: (a) original implementation schema;
(b) new implementation schema. The original number of iteration nbnd are divided
evenly among the band groups used. So each group computes ibnd\_end-ibnd\_start 
iteration. At the ene of the loop the results of each group are summed up.}
\label{fig:gipaw}
\end{figure*}

The development activity focused on three main DFT applications: Nuclear
Magnetic Resonance (NMR), Exact-Exchange (EXX) and Car-Parrinello calculations. Those
computational models are implemented into the Quantum-ESPRESSO distribution 
among three scientific codes named QE-GIPAW~\cite{GIPAW}, PWscf (EXact eXchange part
only) and CP, respectively. 
For all these packages, performance analysis indicated that
highly compute intensive sections were nested into loops over electronic
bands. New levels of parallelism were introduced to better distribute such
computation as these algorithms do not present data dependency along the
band dimension. Applied, this simple concept becomes a breakthrough to
simulate new scientific problems efficiently on large-scale supercomputers
and to discover emerging physical effects that arise at a scale that
was computationally unreachable before. 

One of the most computationally intensive algorithms of the QE-GIPAW code
is to evaluate the linear response of the wave-functions to an external
magnetic field. This is done by solving the Sternheimer equation which
is composed of $n = 1\dots N$ independent linear systems of
the form
\beq 
  \left(\mathcal{H}^{(0)} - E_{n}^{(0)}\right) \ket{\Psi_{n}^{(1)}} = 
  \mathcal{H}^{(1)} \ket{\Psi_{n}^{(0)}}
\eeq
where $\mathcal{H}^{(0)} - E_{n}^{(0)} = \mathcal{G}(E_n)^{-1}$ is the
inverse of the Green's function, $\mathcal{H}^{(1)}$ is the perturbing
magnetic field, and  $\ket{\Psi_{n}^{(0)}}$  are the unperturbed
wave-functions, obtained by a previous SCF calculation with PWscf. Here
$\ket{\Psi_{n}^{(1)}}$  are the unknowns and the index $n$ runs over all
occupied electronic states. Previous to the present work, the Sternheimer
equations were solved band-by-band by the Conjugate Gradient method, with
a clever re-grouping of not converged bands, in order to exploit level-3
BLAS operations. In the new implementation, we further distribute the
occupied bands over CPUs, and the CG algorithm has been modified to
work with groups of bands. The new schema is reported as pseudo-language
shape in Fig.~\ref{fig:gipaw}.  As the numbers of \emph{\emph{electronic bands}}
become relevant the same model might also improves
scalability of plane-wave DFT codes for efficiently runnig at thousands
of cores in parallel.
 
\begin{figure*}[!th]
\centering
\includegraphics[width=5in]{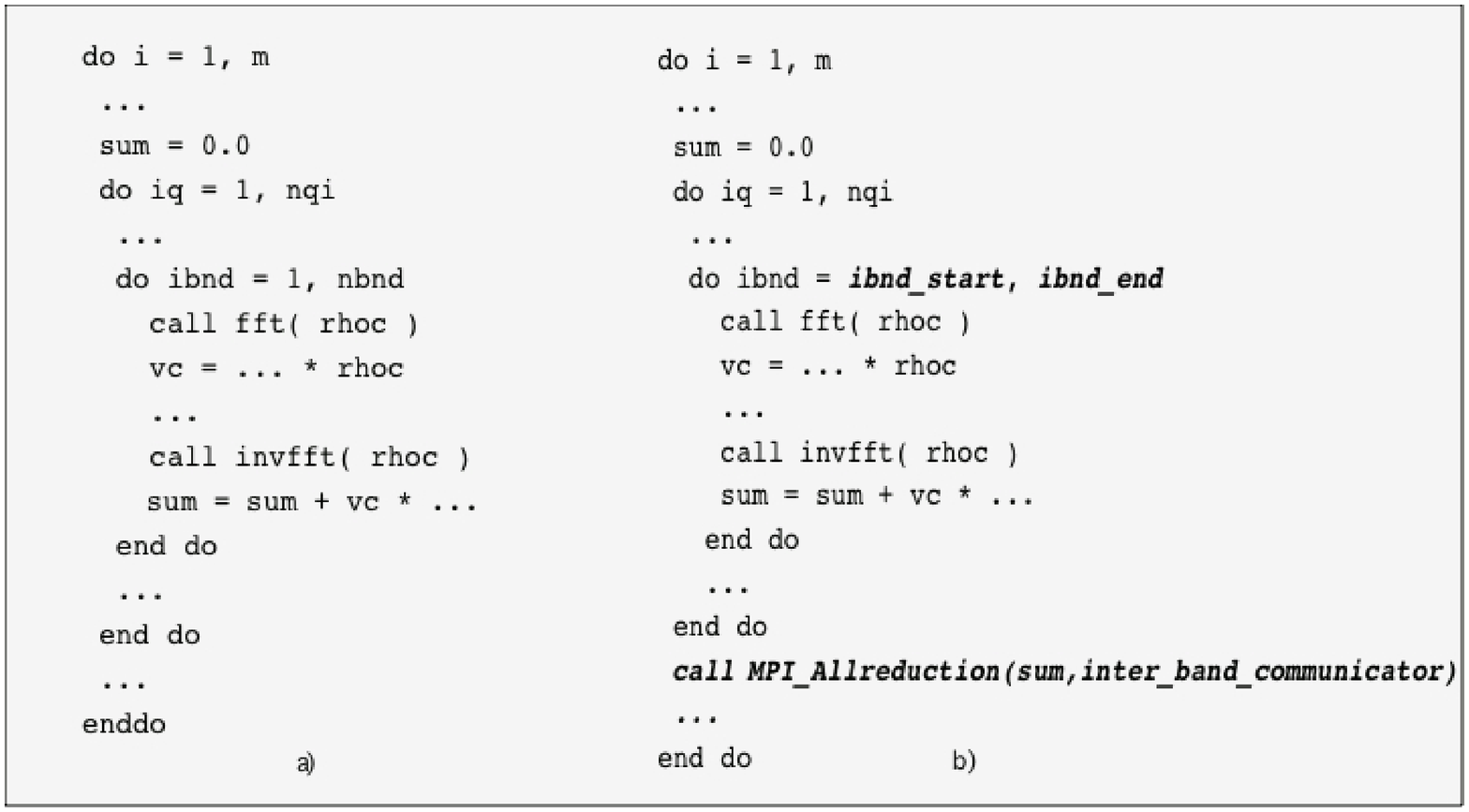}
\caption{Code sections from EXX code: (a) original implementation schema;
(b) new implementation schema. The original number of iteration nbnd are divided
evenly among the band groups used. So each group computes ibnd\_end-ibnd\_start 
iteration. At the ene of the loop the results of each group are summed up.}
\label{fig:exx}
\end{figure*}

The code analysis brought out that the GIPAW calculation is based on a
simultaneous run over a 7-fold loop at one of the outermost routines. The
\emph{image communicator} (see Fig. \ref{fig:parallelization}), 
already implemented in the modular structure
of the package, is designed for such purpose but it was not yet used
for the QE-GIPAW code. A new communicator (later called \emph{electronic
bands} communicator) has been introduced in order to take advantage of
this level of parallelism (see Fig.~\ref{fig:parallelization}). As presented in
the next section even higher scalability is reachable whether this is
exploited. Despite the simple model, the introduction of new levels of
parallelism required a considerable re-factoring due to the complexity of
the software structure.  

A similar parallelization scheme has been
implemented for the Exact-Exchange
(EXX) kernel of the PWscf code. We worked at the bands level as described for
the QE-GIPAW code because the target was to enhance the parallelization for
simulating big systems . As presented in Fig.~\ref{fig:exx} once again it
was possible to parallelize the inner loop over the \emph{\emph{electronic bands}}
by introducing the \emph{\emph{electronic bands}} MPI communicator (i.e.,
by splitting the \emph{ibnd} index across processors to perform on a
different local section of the array \emph{rho}). In this case the parallelization scales much more effectively 
because it is implemented on a high level loop that performs the whole computational workload of
the EXX applications. In other words, the sum over occupied states that
appear in the Green function calculation and in the EXX self-energy has
been distributed across processors. 

The CP code has benefited from a very similar parallelization strategy. 
At large scale, performance analysis underlined that main computational bottlenecks are:  
wave function orthogonalization (the use of  ScaLAPACK could likely be for the better) 
and the evaluation of both forces on electronic degrees of freedom and charge density, 
specially for systems whith a very large number of electrons. 

Both electronic forces and charge density are computed iteratively over the \emph{electronic bands}, performing distributed 3D-FFT at each iteration. 
It is well-known that distributed 3D-FFT's do not scale because those are parallelized along the z axis 
and the number of points in this dimension does not grow at the same
rate as the number of bands. As direct consequence, the overall scalability  is 
limited by parallel FFT up to a few hundreds of tasks. A hybrid MPI+OpenMP parallelization scheme allows to extend the scalability up to few thousands of cores 
(see also CPMD~\cite{hutter05,  hutter06}). 
The band parallelization was introduced to scale the workload of the two loops 
which compute forces and charge density, likewise what we described for QE-GIPAW routines and of the EXX kernel,
see Fig.~\ref{fig:gipaw} and Fig.~\ref{fig:exx}. 
However, contrary to QE-GIPAW and the EXX kernel of PWscf, here, 
data and loops were distributed among the different band groups.
Both data and parallelism over the g-vectors have been replicated, so that all the    
g-vectors loops are left untouched. Although the g-vectors are replicated across band group,
the new implementation requires less memory. The wave functions arrays are distributed along two
dimensions such g-vectors and band index. While two-dimensional arrays having number of atoms and band index as dimensions are distributed on the band index.
Indeed, when the band parallelism is turned off  the memory capacity is a limit for large physical system.

Beside the parallelization of electronic forces and charge density,
the new schema has been partially used in the orthogonalization subroutine too.
In particular band parallelization has been used to distribute the setup of the matrices 
 for the orthogonalization (these matrixes are in fact the
product of one wave function against all the other wave functions),
whereas the diagonalization and the matrix multiplications performed with ScaLAPACK
have been left replicated across the band groups.

In the next section we will show
the scalability improvements obtained with these new parallelization schemes.

\section{Benchmark results}
\label{sec:benchmark}
 Computer technologies are rapidly evolving and the number of compute cores 
 needed to build the most powerful supercomputing infrastructure worldwide 
  is drastically increasing (over a million for the time of writing).  
As a consequence, access to supercomputing petascale facilities commonly calls for
specific requirements, that is scalability. Even for algorithms that are not embarassingly parallel,
speedup can be achived for higher orders of magnitude of core count. However, in such cases,
efficiency will be poor and so less useful as a measure.
In such cases, the need of performing scientific challenges which
would be impossible otherwise, it is a reasonable compromise to balance the
cost of efficiency.  Within this scenario, we enabled three scientific codes
at petascale facilities so that the user community can finally exploit
such compute capabilities. Validation and performance analysis of our
development work was performed using real data. In particular, a
cholesterol system with 592 atoms and 600 \emph{\emph{electronic bands}} was used for
QE-GIPAW tests. The EXX tests have been performed on a 109-atoms
a-SiO$_{2}$ supercell with an oxygen interstitial, while for CP, we have
obtained a series of benchmarks results on a CNT10POR8 system 
(one hydrogen-saturated carbon nanotube, with four porphirin rings 
chemically linked to the CNT surface. The overall system comprise 
1532 atoms, 5232 electrons, i.e, 2616 occupied bands). Calculations have
been run on EU Tier-0 systems: CURIE \cite{Curie}, a 3-fraction linux
cluster power by BULL and equipped with Intel x86-64 based technology
, JUGENE \cite{Jugene} an IBM Blue Gene/P architecture and FERMI
IBM Blue Gene/Q.

\begin{table}\begin{center}
  \begin{tabular}{|l|l|l|l|l|l|}
  \hline
  \# Cores & \# Threads & \# Bands & Time & Efficiency \%\\
  \hline
  64    &     1   &     1    &    621.15   & 100 \\ \hline
  64    &     1   &     2    &    694.66   & 89 \\ \hline
  128   &     1   &     1    &    533.33  & 58 \\ \hline
  128   &     1   &     2    &    416.69  & 73 \\ \hline
  256   &     1   &     4    &    300.79  & 57 \\ \hline
  \end{tabular}
  \caption{GIPAW benchmark data for 592 cholesterol atoms on CURIE~\cite{Curie}.
  Wall-clock times are given in minutes.}
  \label{BenchGIPAW} 
\end{center}\end{table}

\begin{table}\begin{center}
  \begin{tabular}{|l|l|l|l|l|l|}
  \hline
  \# Cores & \# Threads & Time & Efficiency \% & Time Eq.~\ref{eq:sternheimer} \\
  \hline
  128   &     1    &    416.69   & 100 & 242.11  \\ \hline
  896   &     1   &     81.59   & 73 &  58.71  \\ \hline
  1792  &     2    &     51.82   & 58 &  37.60  \\ \hline
  3584  &     2    &     36.00   & 41 &  24.81  \\ \hline
  \end{tabular}
  \caption{GIPAW benchmark data for 592 cholesterol atoms on JUGENE~\cite{Jugene}.
  Wall-clock times are given in minutes. Benchmarks were executed with a number of bands equal 2}
  \label{BenchGIPAW-2} 
\end{center}\end{table}

Tab.~\ref{BenchGIPAW} shows the numbers measured performing GIPAW
calculation on the CURIE Tier-0 system. The first row presents the
reference value obtained running over only two nodes.  Indeed, 64 cores
is the lowest value that would allow to perform such calculation within
the 24-hours of wall clock limit and it is necessary to reach convergence
to obtain timing information.  The second row reports the elapsed time we
got with the new version using two band groups, 32 cores for each group.
Here the old approach still performs better since the diagonalization
routine is not parallelized over the bands. However, by doubling the
number of cores the parallelization over the \emph{\emph{electronic bands}} reduced
the wall-time.  The fourth row of Tab.~\ref{BenchGIPAW} presents in fact
better lower value of execution time and consequently a better efficiency
(5th column). De facto, 128 cores was the limit of the older version of
the code as already at this stage efficiency is low.  We obtain the
same value of efficiency running the new version of the code at 256 cores using
4 bands group. The new level of parallelism introduced allows at large number of cores
to increase the scalability by a factor of two. While the Green's function
scales almost perfectly the code sections the iterative diagonalization
is now the actual bottleneck. At this level 1/3 of the overall time is spent
to solve the diagonalization problem that does not take advantage from the
parallelization over \emph{\emph{electronic bands}}. The parallelization over electronic
bands cannot then be further applied to improve the scalability. So,
QE-GIPAW runs efficiently with 2 band groups for this example. This
is the limit where users are suppose to utilize the other level
of parallelism introduced. 

As described in the previous section the code
implements seven independent calculations at the outermost routine. For
our experiments we took the limit at 128 cores. As shown in
Tab.~\ref{BenchGIPAW-2}, by using this new parallelism and running over
896 cores the code scales with 73\% of efficiency respect to 128 cores
(93\% if compared with the same number of cores while using the old
version of the code). Although this is the best result reached in term
of efficiency, further scaling with the introduction of hybrid MPI+OpenMP
approach shows that is possible to reduce up to 36 minutes 
 while running at 3584 cores. Thanks to the new development we
present in this paper users can now easily scale at least 14 times the
number of cores in regards to the old version of the same code.

The result analysis continues now with the EXX calculation. As described
in the previous section the routine \emph{Vexx} is carrying out almost
all of the computational workload. In order to test the impact of the new
parallelization we simulate a system of 108 atoms with 800 bands on the
Tier-0 JUGENE machine. Despite the low number of atoms, full convergence
requires a huge amount of core hours as shown in Tab.~\ref{BenchEXX},
almost 300,000 hours were necessary to complete this computation. For
 completeness it must be underlined that the current implementation
allows only norm-conserving pseudopotentials~\cite{TM} and by consequence
the energy cutoff used for the fock operator (EXX operator) is the
density cutoff, i.e. four times the wave function cutoff, although a
smaller cutoff can be used.

\begin{table}\begin{center}
  \begin{tabular}{|l|l|l|l|l|}
  \hline
  \# Cores & \# Bands & Elapsed Time(s) & Efficiency \% \\
  \hline
  32768   &   64   &   523.45 & 100  \\ \hline
  65536   &   128    &   311.59 & 84 \\ \hline
  \end{tabular}
  \caption{EXX benchmark data for 109 SiO$_2$ atoms on JUGENE~\cite{Jugene}.}
  \label{BenchEXX}
\end{center}\end{table}

The routine \emph{Vexx}, that calculates the Exact-Exchange potential,
scales perfectly by doubling the number of cores and the number of
band groups. We measure an efficiency of around $\sim$97\% moving from
32768 to 65536 cores if we compare the time of execution of the single
Vexx routine. However, as presented in Tab.~\ref{BenchEXX} also the
global time scales almost perfectly at large number of cores. Indeed,
since \emph{Vexx} is extremely computationally intensive in regards to
the overall time of execution, the code scales efficiently up to 65536
cores. The level of parallelism introduced by the new \emph{elctronic
bands} communicator definitely impacts more the EXX code than the QE-GIPAW
code as in this case almost all of the computational workload can be
parallelized over the electronics bands.

\begin{figure}
\includegraphics[width=3.5in]{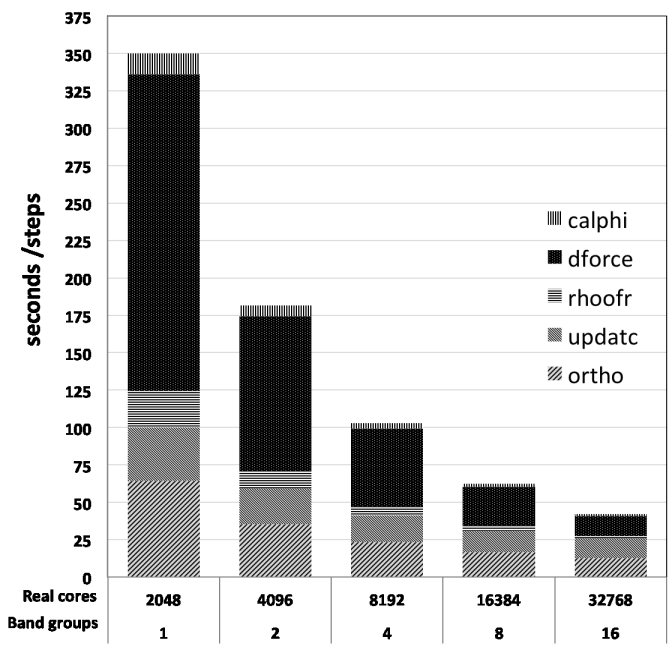}
\caption{Scalability of the CP kernel of Quantum ESPRESSO on BGQ, using the benchmark 
CNT10POR8. A 1532 atoms system of an hydrogen-saturared chiral carbon nanotube, 
with four phorphyrin rings chemically linked to the surface of the nanotube}
\label{fig:CP-BGQ}
\end{figure}

The benchmark analysis for the CP code, see Fig.~\ref{fig:CP-BGQ}, shows that,
 the time spent in the ortho subroutines
decreases when increasing the number of cores, but not linearly,
whereas the computation of dforce and rhoofr as well as other
subroutine containing loops over \emph{electronic bands} scale almost linerly with
the number of cores.
It is worth to note that, when increasing the number of band groups 
to the limit in which electronic force and charge density computation become negligible,
the linear algebra computation contained in the ortho subroutine
will become again the main bottleneck of the code.
At that point some new strategy has to be implemented.

\section{Conclusions}
\label{sec:conclusion}
The work presented in this paper shows the
advantage of the band parallelization approach to simulate challenging
systems when the number of bands become huge. This was implemented for two DFT applications.
(In fact, in order to accurately reproduce experimental data it is often necessary to keep the
system size big enough). The results presented here suggest that it
is mandatory to use this strategy on petascale hardware and beyond.
Recently the memory distribution in some critical parts of Quantum-Espresso has been
introduced. So, those two strategies coupled together are the workhorse
that allow efficient simulation of DFT calculation on world-class
supercomputers.

\section*{Acknowledgements}
This work was financially supported both by the PRACE FirstImplementation Project funded in 
part by the EUs 7th Framework Programme (FP7/2007-2013) under grant agreement no. RI-261557 
and by Science Foundation Ireland (grant 08/HEC/I1450).
Benchmarks were carried out on the Jugene~\cite{Jugene} machine at Jülich, on
Curie~\cite{Curie} at the CEA and on Fermi\cite{Fermi}.  DC is in indebted with Ari P. Seitsonen,
Uwe Gerstmann and Francesco Mauri for coauthoring the first version
of the QE-GIPAW code. We acknoweledge Emine Küçükbenli for sharing the
cholesterol model, Arrigo Calzolari for the CNT10POR8 model, and
Paolo Giannozzi for useful discussions. We acknowledge the Quantum ESPRESSO Foundation
for the support.



\end{document}

%% file: macros.tex
\newcommand{\beq}{\begin{equation}}
\newcommand{\eeq}{\end{equation}}
\newcommand{\beqa}{\begin{eqnarray}}
\newcommand{\eeqa}{\end{eqnarray}}
\newcommand{\bfig}{\begin{figure}\begin{center}}
\newcommand{\efig}{\end{center}\end{figure}}
\newcommand{\btab}{\begin{table}\begin{center}}
\newcommand{\etab}{\end{center}\end{table}}

\newcommand{\Hcal}{\ensuremath{\mathcal{H}}}

\newcommand{\Ocal}{\ensuremath{\mathcal{O}}}
\newcommand{\Gcal}{\ensuremath{\mathcal{G}}}

\newcommand{\rvec}{\ensuremath{\mathbf{r}}}
\newcommand{\Rvec}{\ensuremath{\mathbf{R}}}
\newcommand{\vvec}{\ensuremath{\mathbf{v}}}

\newcommand{\Bvec}{\ensuremath{\mathbf{B}}}

\newcommand{\Gvec}{\ensuremath{\mathbf{G}}}
\newcommand{\jvec}{\ensuremath{\mathbf{j}}}
\newcommand{\Jvec}{\ensuremath{\mathbf{J}}}
\newcommand{\kvec}{\ensuremath{\mathbf{k}}}

\newcommand{\qvec}{\ensuremath{\mathbf{q}}}

\newcommand{\nn}{\nonumber\\}
\newcommand{\tensor}[1]{\ensuremath{\overleftrightarrow{#1}}}